\documentclass[%
reprint,
amsmath,amssymb,
aps,
]{revtex4-1}

\usepackage{graphicx}
\usepackage{dcolumn}
\usepackage{bm}
\usepackage{xcolor}

\newcommand{\E}{\varepsilon}
\newcommand{\A}{\alpha}
\newcommand{\B}{\beta}

\renewcommand{\bf}{\textbf}

\begin{document}
	
	\preprint{APS/123-QED}
	
	\title{Controlling diffraction patterns with metagratings}

		\author{Vladislav Popov}
	\email{uladzislau.papou@centralesupelec.fr}
	\affiliation{%
		SONDRA, CentraleSup\'elec, Universit\'e Paris-Saclay, F-91190, Gif-sur-Yvette, France
	}%
	\author{Fabrice Boust}%
	\email{fabrice.boust@onera.fr}
	\affiliation{%
		SONDRA, CentraleSup\'elec, Universit\'e Paris-Saclay,
		F-91190, Gif-sur-Yvette, France
	}
	\affiliation{%
		DEMR, ONERA, Universit\'e Paris-Saclay, F-91123, Palaiseau, France
	}%
	
	\author{Shah Nawaz Burokur}%
	\email{sburokur@parisnanterre.fr}
	\affiliation{%
		LEME, UPL, Univ Paris Nanterre, F92410, Ville d'Avray, France
	}%

\begin{abstract}
In this study we elaborate on the recent concept of metagratings  proposed in Ra'di \textit{et al.} [Phys. Rev. Lett. 119, 067404 (2017)] for efficient manipulation of reflected waves.
Basically, a metagrating is a set of 1D arrays of polarization line currents which are engineered to cancel scattering in undesirable diffraction orders.
We consider a general case of metagratings composed of $N$ polarization electric line currents per supercell. This generalization is a necessary step to totally control diffraction patterns.
We show that a metagrating  having $N$ equal to the number of plane waves scattered in the far-field can be used for controlling the diffraction pattern.
To validate the developed theoretical approach,  anomalous  and  multichannel reflections are demonstrated with 3D full-wave simulations in the microwave regime at $10$ GHz.
The results can be interesting for the metamaterials community as allow one to significantly decrease the number of used elements and simplify the design of wavefront manipulation devices, what is very convenient for optical and infra-red frequency ranges. Our findings also may serve as a way for development of efficient tunable antennas in the microwave domain.
\end{abstract}

\maketitle

    For long time, the microwave community has approached a particular problem of anomalous reflections  by means of reflectarray antennas~\cite{Hum_2005,Hum_2007}.
     In such antennas, a linear phase variation is created along the surface, allowing one to reflect incident waves to a desirable angle. 
With the development of nanofabrication technologies and metasurfaces, the concept of reflectarrays was transposed to infra-red and optical frequency domains~\cite{Capasso_GeneralizedReflectionLaw,Glybovski2016}.
A metasurface is represented by a 2D dense distribution of subwavelength scatterers and a reflectarray is a particular case of a metasurface which can generally be used for various applications other than anomalous reflections.
However, reflectarrays suffer from low efficiencies for angles of anomalous reflection approximately greater than $45$ degrees~\cite{Asadchy2016_SpatiallyDispMS}.
Stimulated by the advances in fabrication technologies, extensive research in the area established a strong theoretical ground in the form of equivalence principle~\cite{Pfeiffer2013} for the design of wavefront manipulation devices based on the use of metasurfaces. As such, multichannel reflection with metasurfaces was demonstrated both theoretically and experimentally in~\cite{PhysRevX.7.031046}. 
Recently, a metasurface performing highly efficient anomalous reflection at steep angle has been demonstrated in~\cite{Tretyakov2017_perfectAR} on the basis of the concept of  metasurfaces possessing strong spatial dispersion~\cite{Asadchy2016_SpatiallyDispMS,Epstein2016_AuxiliryFields}.

\begin{figure}[tb]
\includegraphics[width=0.99\linewidth]{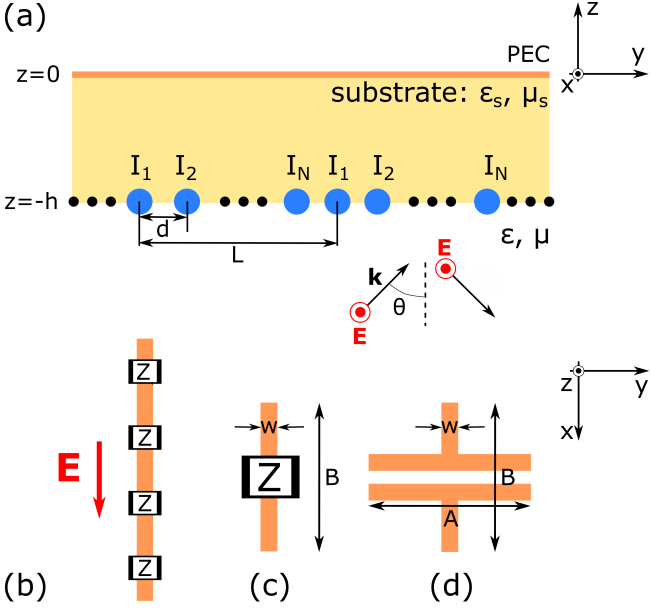}
\caption{\label{fig:1} (a) System under consideration: a periodic array of line currents $\bf J_{nq}=I_q\exp[-jk\sin[\theta]nL]\delta(y-y_{nq},z+h)\bf x_0$ (blue circles) placed on PEC-backed dielectric substrate having permittivity $\E_s$, permeability $\mu_s$ and thickness $h$. The array is excited by a plane wave  incident at angle $\theta$ and having  TE polarization. (b) A line current implemented as a 1D array of loaded dipoles. (c) A PEC strip dipole of length $B$ and width $w$ loaded with lumped impedance $Z$. (d) A PEC strip dipole loaded with printed circuit capacitance having arms of length $A$.}
\end{figure}

Unfortunately, a theoretical framework to design strongly spatial dispersive metasurfaces has not been developed yet, making the design of a sample time consuming [if it is possible at all] as it requires 3D full-wave simulations. 
In spite of advances in the field of metasurfaces,  drawbacks concerning design complexity and material losses still exist, rendering implementation of high performance devices very challenging in some frequency ranges~\cite{Ratni_18}.

In this study we elaborate on the recent concept of metagratings~\cite{Alu2017_single_UC} for the manipulation of reflected waves.
Basically, a metagrating is a set of 1D arrays of scatterers such as polarization line currents, separated by a distance of the order of the operating wavelength $\lambda$.
Polarization line currents are used to cancel scattering in undesirable diffraction orders.
Metagratings allow one to significantly decrease the number of constitutive scatterers in contrast to metasurfaces where scatterers are tightly packed in the plane.
This reduction can be  very attractive to reduce the fabrication complexity as well as joule losses.

On the theoretical level, metasurface and metagrating are described differently. As a metasurface is composed of deeply subwavelength tightly packed elements, one can introduce averaged surface impedances. Meanwhile, a metagrating is treated as an array of polarization line currents separated by distances much larger than their sizes. Even though, there can be many polarization line currents in a supercell the separation between the currents  remains on the order of operating wavelength  and one would speculate by introducing average surface impedances.

It has been already shown that having just a single line current per period allows one to cancel specular reflection and perform perfect beam splitting and anomalous reflection~\cite{Alu2017_single_UC,Epstein2017,Epstein2018}.
In~\cite{PhysRevX.8.011036}, the authors numerically and experimentally demonstrated the possibility to perform highly efficient broadband anomalous reflection with a Huygens' metasurface having just two meta-atoms per supercell necessary for cancelling specular reflection. 
Basically, the same functionality was demonstrated in~\cite{Alu2017_single_UC,Epstein2018} where  a single meta-atom per supercell and the substrate thickness were used as degrees of freedom instead of two meta-atoms per super cell. In this sense, the work in Ref.~\cite{PhysRevX.8.011036} is very similar to the ones on metagratings.
Chalabi \textit{et al.} also demonstrated the possibility to perform near-perfect anomalous reflection using two line currents per super cell~\cite{Alu2017_two_UCs} that are necessary for eliminating reflection in the zeroth and minus first diffraction orders. 
Recently, an implementation of a graphene-based tunable metagrating operating in the THz frequency range was suggested in~\cite{Alu_two_UC_reconfig}.

In the present work, we study a general case of metagratings having $N$ polarization line currents per super cell. 
This generalization is a necessary step for controlling  diffraction patterns when the number of plane waves scattered in the far-field is greater than three.
Although, the authors of Ref.~\cite{PhysRevX.8.011036} discussed the number of meta-atoms per super cell necessary for controlling arbitrary number of plane waves diffracted in the far-field, a clear theory for designing a N-meta-atoms Huygens' metasurface was not elaborated.

Gaining control over many diffraction orders can be particularly interesting for implementing tunable devices and performing multichannel reflection. Indeed, having many identical wires but being able to control polarization currents in each of them allows one to perform all possible transformations  of the  diffraction pattern with the same device, where the only restriction remains the device size.
Moreover, metagratings can operate in a broad frequency range as usually does not require  resonance response of meta-atoms. Broadband response of metagratings with a single and couple of polarization currents per supercell was demonstrated in~\cite{Epstein2018} and~\cite{Alu2017_two_UCs}, respectively.

\begin{figure}[tb]
\includegraphics[width=0.99\linewidth]{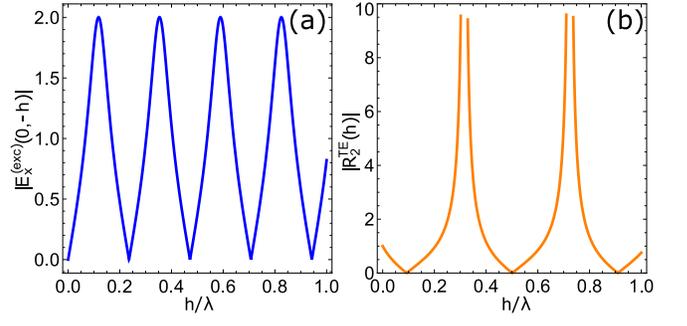}
\caption{\label{fig:4} (a) Dependence of the excitation field [$\theta=0$] acting on a metagrating on the thickness of the substrate $h$ when $\E_s=4.5$ and $\mu_s=1$. (b) Absolute values of the $R_2^{TE}$ vs. the thickness of the substrate when $\theta=0$, $r=l=1$, $N=3$ and $L=\lambda/\sin[60^o]$. The rest of $R_m^{TE}$ does not have poles under these parameters. $\lambda$ is the operating vacuum wavelength.}
\end{figure}

As physical system, we consider a 1D periodic array of polarization electric line currents  placed over a grounded dielectric substrate of thickness $h$  and excited by an incident harmonic TE-polarized plane wave at angle $\theta$ where $\exp[j\omega t]$ time dependence is assumed. 
The array has period $L$ and consists of  super cells  each having  $N$ equally separated line currents by the distance $d=L/N$. 
The schematics of the considered system is presented in Fig.~\ref{fig:1} (a).  
A line current is imagined as a tightly packed row of point dipoles orientated in the same direction, see Fig.~\ref{fig:1} (b). 
Practically, one can realize the dipoles as the loaded rods considered in Fig.~\ref{fig:1} (c) and (d).

\begin{figure*}[tb]
\includegraphics[width=0.99\linewidth]{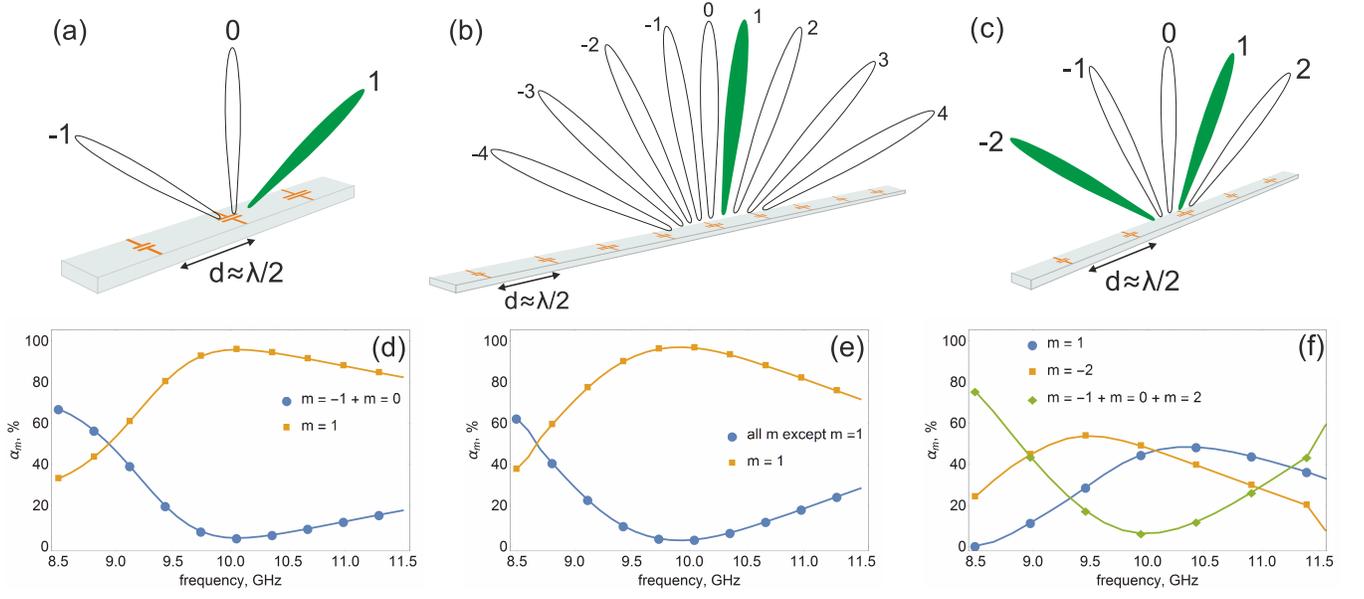}
\caption{\label{fig:5} The top row of figures demonstrates schematics of simulated metagratings with (a) $N=3$, (b) $N=9$ and (c) $N=5$, the green and white lobes depict excited and canceled diffraction orders respectively. Figures in the bottom row depict obtained from 3D full-wave simulations frequency responses of the metagratings corresponding to the figures in the top row.
(a), (d) The example of anomalous reflection at angle of $50^o$ with the metagrating having $N=3$, $L=\lambda/\sin[50^o]$, and loads [in $\eta/\lambda$, $\tilde{\kappa}=26.80$ mm]  $Z_1=-14.5j, Z_2=-6.86j, Z_3=-4.43j$.
%
(b), (e) The example of small-angle anomalous reflection [of $12.5^o$] with metasurface  having $N=9$,  $L=4\lambda/\sin[60^o]$ and loads [in $\eta/\lambda$,  $\tilde{\kappa}=27.43$ mm]  $Z_1=-j7.62, Z_2=-j6.96, Z_3=-j6.19, Z_4=-j5.55, Z_5=-j5.18, Z_6=-j3.57, Z_7=-j3.02, Z_8=-j18.7, Z_9=-j10.1$. 
(c), (f) The example when out of five only $-2^\textup{nd}$ and $1^\textup{st}$ diffraction orders are equally excited with the metagrating having $N=5$, $L=2\lambda/\sin[50^o]$ and loads [in $\eta/\lambda$, $\tilde{\kappa}=27.18$ mm] $Z_1=-9.00j, Z_2=-5.88j, Z_3=-6.59j, Z_4=-3.03j, Z_5=-5.14j$. 
The substrate is Arlon AD450 [$\E_s=4.5$], $h=3$ mm, $B=\lambda/10=3$ mm and $w=3\textup{mil}\approx 76.2\mu\textup{m}$.
%
%
}
\end{figure*}

In the presence of the grounded substrate the excitation field takes the following form
\begin{eqnarray}\label{eq:Eexc_TE}
&&E^{exc}_x(y,z\leq-h)=\left(e^{-j\B_0z}+R_0^{TE}e^{j\B_0(z+2h)}\right)e^{-jk\sin[\theta]y}.\nonumber\\
\end{eqnarray}
Electric line currents in the array are represented as current densities $\bf J_{nq}(\bf r)=I_q\exp[-jk\sin[\theta]nL]\delta(y-y_{nq},z+h)\bf x_0$ where $\delta(y,z)$ is the Dirac delta function, $y_{nq}= nL+(q-1)d$, $n$ and $q$ take integer values from $-\infty$ to $+\infty$ and from $1$ to $N$, respectively. The term $\exp[-jk\sin[\theta]nL]$ represents the phase variation of the currents introduced by the incident wave. Radiation of the array  of electric line currents is represented by a series of Hankel functions~\cite{felsen1994radiation,tretyakov2003analytical} of the second kind.
It can be shown by means of the Poisson's formula [see Supplementary Materials] that the electric field of the wave radiated by the array outside the substrate can be written as 
\begin{eqnarray}\label{eq:ELC_TE_sub}
&&E_x(y,z<-h)=\nonumber\\
&&-\frac{k\eta}{2L}\sum_{m=-\infty}^{+\infty}\frac{\rho^{(I)}_m(1+R_m^{TE})}{\B_m}e^{-j\xi_my+j\B_m(z+h)},
\end{eqnarray}
$E_y=E_z=0$. Corresponding magnetic fields can be found from  the Maxwell equations.
The series represent superpositions of plane waves having tangential component of wave vector equal to $\xi_m=k\sin(\theta)+2\pi m/L$, the longitudinal component is given by $\B_m=\sqrt{k^2-\xi_m^2}$ outside  and by $\B_m^s=\sqrt{k_s^2-\xi_m^2}$ inside the substrate [$k=\omega\sqrt{\E\mu}$ and $k_s=\omega \sqrt{\E_s\mu_s}$ are respectively the wave numbers outside and inside the substrate].
Thus, $R_m^{TE}$ is Fresnel's reflection  coefficient from the grounded substrate of a plane wave having tangential component of the wave vector equal to $\xi_m$.
Each current contributes to the amplitudes of the plane waves via the introduced quantity $\rho^{(I)}_m$
\begin{equation}\label{eq:rho_TE}
\rho^{(I)}_m=\sum_{q=1}^{N}I_q\exp[j\xi_m (q-1)d].
\end{equation}
One can recognize in Eq.~\eqref{eq:rho_TE} a discrete Fourier transformation.

In general case when a plane wave illuminates a metagrating one can find $r+l+1$ scattered plane waves in the far-field, where $r$ and $l$ are  largest integers satisfying the conditions $\B_{r}>0$ and $\B_{-l}>0$.
However, we can arbitrary control all of the $r+l+1$ plane waves if the number $N$ of line currents  in a super cell  is equal to $r+l+1$. 
Indeed, amplitude $A_m^{TE}$ of the $m^{th}$  plane wave depends on $\rho_m^{(I)}$ which is determined by the currents $I_q$ [see Eq.~\eqref{eq:ELC_TE_sub}]
\begin{equation}\label{eq:A}
A_m^{TE}=-\frac{k\eta}{2L}\frac{\rho^{(I)}_m
(1+R_m^{TE})e^{j\B_m h}}{\B_m}+\delta_{m0}R_0^{TE}e^{2j\B_0h},
\end{equation}
where $\delta_{m0}$ is the Kronecker delta accounting for the incident wave reflected from the substrate.
By setting all the amplitudes $A_m^{TE}$ ($m\in[-l,r]$), one can find necessary $I_q$ from the corresponding $\rho_m^{(I)}$ which are related via Eq.~\eqref{eq:rho_TE}.

Thus, by designing currents $I_q$ one can perform all possible transformations of the  diffraction pattern, e.g. beam splitting, anomalous reflection, multichannel reflection, etc.
When implementing line currents as thin perfectly conducting wires one can obtain necessary currents $I_q$ by loading wires with suitable impedance densities $Z_q$. The last are found from  
\begin{equation}\label{eq:Z}
Z_qI_q=E_x^{(exc)}(y_{0q},-h)-Z_{in}I_q-\sum_{p=1}^N Z_{qp}^{(m)}I_p
\end{equation}
where the right-hand side simply represents total electric field at the location of the $q$th wire  in the zeroth supercell [$y_{0q}=(q-1)d$, $z=-h$]. Here, we also introduce notations for the input impedance density of  wires $Z_{in}=k\eta H_0^{(2)}[kr_0]/4$, with $H_0^{(2)}[kr_0]$ being the Hankel function of the second kind and $r_0$ the radius of the wires, and for the mutual impedance densities $Z_{qp}^{(m)}$ which account for interaction between the wires and between the wires and the  grounded substrate. When a wire is realized as a perfectly conducting strip of width $w$, the radius is $r_0=w/4$~\cite{tretyakov2003analytical}. Derivation of Eq.~\eqref{eq:Z} and expressions for the mutual impedance densities can be found in Supplementary Materials.

Generally, currents found from~\eqref{eq:A} correspond to active and lossy loads $Z_q$ calculated from~\eqref{eq:Z}.
From a practical point of view, we are interested only in passive and lossless metagratings where $\Re[Z_q]=0$, i.e. which cannot radiate energy by themselves and do not require engineered joule losses. 
A metagrating should redistribute the energy of the incident wave between $r+l+1$ diffracted in the far-field plane waves. Then, the power conservation condition when assuming a unity amplitude of the incident wave reads as 
\begin{equation}
\label{eq:power_conserv}
\sum_{m=-l}^{r}\A_m=1,\quad \A_m=\left|A_m^{TE}\right|^2\frac{\B_m}{\B_0},
\end{equation}
where $\A_m$ is the part of the incident energy going in the $m^{th}$ diffraction order.

In contrast to  the case of metagratings having a single line current in a super cell~\cite{Alu2017_single_UC,Epstein2017,Epstein2018}, when it comes to greater number of line currents per supercell there are no exact analytical formulas for reactive load impedance densities necessary for obtaining some diffraction pattern.
To approach this problem we develop a very simple real valued genetic algorithm~\cite{WRIGHT_1991} which allows one to find reactive  $Z_q$ with given impedance reactivity accuracy $p$ for a desired  diffraction pattern obtained with given transformation accuracy $\A$. The impedance reactivity accuracy is defined in accordance with the following inequality $\sqrt{\sum_{q=1}^{N}|Re[Z_q]/Z_q|^2}<p.$
A diffraction pattern is set by assigning to all $\A_m^0$ certain values. Phases $\phi_m=\arg[A_m^{TE}]$ are assumed to be not important and assigned randomly. Transformation accuracy $\A$ means that one is satisfied with a transformation when the part of the incident energy going in the $m^{th}$ diffraction order is within the range $\A_m=\A_m^0\pm\A$. 
Still, at each step the genetic algorithm deals with $\A_m>0$ constrained by the energy conservation condition~\eqref{eq:power_conserv}.


When designing a metagrating, one should also take care of choosing parameters of the substrate.
First of all, when substrate's thickness is varied the value of the excitation field  Eq.~\eqref{eq:Eexc_TE} on a metagrating passes through zeros as illustrated in Fig.~\ref{fig:4} (a). Clearly, a metagrating cannot be excited when the excitation field is zero on its plane.
And secondly, the reflection  coefficient $R_m^{TE}$ as a function of $h$ has poles when $m$ is such that $k<\xi_m$ but $k_s>\xi_m$ [Fig.~\ref{fig:4} (b)]. The poles correspond to excitation of  waveguide modes inside the substrate. 
Thus, assuming $\E_s$ and $\mu_s$ of the substrate are set, one should choose the thickness: (i) corresponding to vicinity of the maximum of the excitation field on a metagrating and (ii) $|R_m^{TE}(h)|\neq\infty$.

 One can  realize the line currents as dense 1D arrays of loaded dipoles [separated by distance $B\ll\lambda$ and having lumped load equal to $Z$] as in Figs.~\ref{fig:1} (b) and (c). 
Then, the load impedance density is simply $Z/B$. A capacitive load can be realized as a printed circuit capacitance as illustrated in Fig.~\ref{fig:1} (d) for which $Z=-j \eta \kappa/(A\E_{eff})$ [when other parameters $B$ and $w$ are fixed], $\kappa$ is the proportionality factor, $\E_{eff}$ is approximated as $(1+\E_s)/2$ and $\mu_s$ is assumed equal to $1$.
The proportionality factor $\kappa$  was introduced in~\cite{Epstein2017_metagrating} for the case of a single line current per supercell. However, it turns out that in the general case of many line currents per supercell one can successfully use the same proportionality factor for all  currents, i.e. independently on $q$. Thus, when load impedance densities $Z_q$ are found from the genetic algorithm, one can easily calculate arms lengths of necessary printed capacitors as $A_q=-\tilde{\kappa}/(\Im[Z_q]\frac{\lambda}\eta \E_{eff})$, $\tilde{\kappa}=\lambda\kappa/B$.

In order to validate the developed theoretical basis, we perform 3D full-wave simulations  with \textit{COMSOL Multiphysics}.
We demonstrate three examples of metagratings designed to operate at $10$ GHz [$\lambda\approx 30$ mm] and perform different transformations of the diffraction pattern as shown in Fig.~\ref{fig:5}.
A polarization line current is implemented as a 1D array of capacitively loaded  perfectly conducting strips [as it schematically shown in the top row of Fig.~\ref{fig:5}].
In all the examples normally incident  plane wave is assumed, i.e. $\theta=0$.

When performing a large-angle anomalous reflection with a metagrating, overall there are three diffraction orders and therefore only three polarization line currents per supercell are necessary to cancel the $-1^\textup{st}$ and $0^\textup{th}$ diffraction orders, as illustrated in Fig.~\ref{fig:5} (a).  Figure~\ref{fig:5} (d) depicts the frequency response of the metagrating performing anomalous reflection at angle of $50^o$. The situation is more difficult in case of a small-angle anomalous reflection with the presence of many high diffraction orders and when the energy should be scattered only in the first one. Indeed, in the example of Fig.~\ref{fig:5} (b) there are nine diffraction orders and the metagrating with  nine polarization currents is used to cancel scattering in all of them except the first one corresponding to an anomalously reflected wave at angle of $12.5^o$. Figure~\ref{fig:5} (e) demonstrates the frequency dependence of the metagrating's performance efficiency.
Clearly, metagratings are not restricted to anomalous reflection application and can be used for multichannel reflection. One can distribute the energy of an incident wave between all diffraction orders in a desirable manner. For instance, Figs.~\ref{fig:5} (c) and (f) demonstrate the scenario when the metagrating having five polarization currents is used to  split normally incident waves between the $-2^\textup{nd}$ and $1^\textup{st}$ diffraction orders and cancel scattering in the other three diffraction orders.

In conclusion, it has been shown that a metagrating  having the number of polarization line currents per super cell equal to the number of plane waves scattered in the far-field can be used for controlling the diffraction pattern.
Namely, equations~\eqref{eq:rho_TE} and \eqref{eq:A} allowing one to find  currents realizing desirable transformations have been derived. Since there are no analytical formulas of reactive load impedance densities~\eqref{eq:Z} for direct design, genetic algorithms have been implemented for that purpose.
The diffraction orders control has been demonstrated by means of 3D full-wave simulations on the examples of anomalous reflection  and equal redistribution of the energy of the incident wave between two diffraction orders. 

The validation results can be very interesting for the metamaterials community to perform highly efficient control of light scattering. It allows one to significantly decrease the number of used elements and simplify the design, which is very convenient for optical and infra-red frequency ranges. Our findings also may serve as a way for development of efficient tunable antennas in the microwave domain.

\bibliography{bib}

\end{document}